\documentclass[reprint, superscriptaddress, amsmath,amssymb, aps]{revtex4-1}
\usepackage{graphics}
\usepackage{graphicx}
\usepackage{subfigure}
\usepackage{dcolumn}
\usepackage{bm}
\usepackage{hyperref}
\usepackage{color}

\begin{document}
\title{Dual-polarization Second-Order Photonic Topological Insulators}
\author{Yafeng Chen}
 \email{These authors contribute equally to this work.}
\affiliation{Key Laboratory of Advanced Technology for Vehicle Body Design and Manufacture, Hunan University, Changsha, 410082, China}
\author{Fei Meng}
 \email{These authors contribute equally to this work.}
\affiliation{Faculty of Science, Engineering and Technology, Swinburne University of Technology, Hawthorn, VIC 3122, Australia}
\author{Zhihao Lan}
 \email{z.lan@ucl.ac.uk}
\affiliation{Department of Electronic and Electrical Engineering, University College London,
Torrington Place, London, WC1E 7JE, United Kingdom}
\author{Baohua Jia}
 \email{bjiag@swin.edu.au}
\affiliation{Faculty of Science, Engineering and Technology, Swinburne University of Technology, Hawthorn, VIC 3122, Australia}
\author{Xiaodong Huang}
 \email{xhuang@swin.edu.au}
\affiliation{Faculty of Science, Engineering and Technology, Swinburne University of Technology, Hawthorn, VIC 3122, Australia}

\date{\today}

\begin{abstract}
Second-order photonic topological insulators that host highly localized corner states resilient to defects, are opening new routes towards developing fascinating photonic devices. However, the existing works on second-order photonic topological insulators have mainly focused on either transverse magnetic or transverse electric modes. In this paper, we propose a dual-polarization topological photonic crystal structure for both transverse magnetic and transverse electric modes through topology optimization. Simple tight-binding lattice models are constructed to reveal the topological features of the optimized photonic crystal structure in a transparent way. The optimized dual-polarization second-order photonic topological insulator hosts four groups of corner states with different profiles and eigenfrequencies for both the transverse magnetic and transverse electric modes. Moreover, the robustness of theses corner states against defects is explicitly demonstrated. Our results offer opportunities for developing polarization-independent topological photonic devices.

\end{abstract}

\maketitle

\section{\label{sec:intro}Introduction}
One of the most important recent developments of condensed-matter physics is the discovery of topological states of matter \cite{Thouless82PRL,  Klitzing80PRL, Haldane88PRL, KaneMele05PRL_QSH, XueZhang13Science}, in particular, topological insulators (TIs) \cite{TI_review1, TI_review2}. A unique feature of these systems is the existence of topological protected edge states, where according to the bulk-edge correspondence principle, a d-dimensional TI supports (d-1) dimensional gapless edge states. Inspired by the interesting physics and promising technological applications, the concept of TIs has been extended to classical wave systems, e.g., photonic systems \cite{ShvetsNP17review,review_RMP19,QHE_PRL08_Haldane, QHE_Nature09_Wang, WuHu15PRL,Yang18PRLexp, Menglin_TAP, Zhu18PRBdeform, Jiang16OEcoreshell, Chen19PSSRRL}. Photonic topological insulators enable the robust manipulation of light, thus providing the possibility of novel topological photonic applications, such as non-reciprocal devices \cite{QHE_Nature09_Wang}, pseudospin-polarized waveguides \cite{WuHu15PRL, Yang18PRLexp, Menglin_TAP} and topological lasers \cite{Bahari17Science, Harari18ScienceTItheory, TIlaser_Science18Exp}.

Beyond the traditional bulk-boundary correspondence, high-order TIs can host lower-dimensional boundary states \cite{HTI_SciAdv18,benalcazar2017quantized,benalcazar2017electric}. For example, a d-dimensional second-order TI supports (d-1) dimensional {\it gapped} boundary states but (d-2) dimensional {\it gapless} edge states, which are corner states in two-dimensional (2D) systems. The second-order photonic TIs (SPTIs) have been theoretically predicted and experimentally demonstrated \cite{Xie18PRBcorner, Chen19PRLhighorderTI, Dong19PRLsecondorderTI, Hassan19NPcorner, Chen20PRRinverse,kim2020multipolar,mittal2019photonic}. Based on these highly localized corner states, high Q factor nanocavity \cite{Ota19Optica} and low-threshold topological nanolaser \cite{Zhang20LightCorner} have been experimentally realized.  

\begin{figure}
\includegraphics[width=\columnwidth]{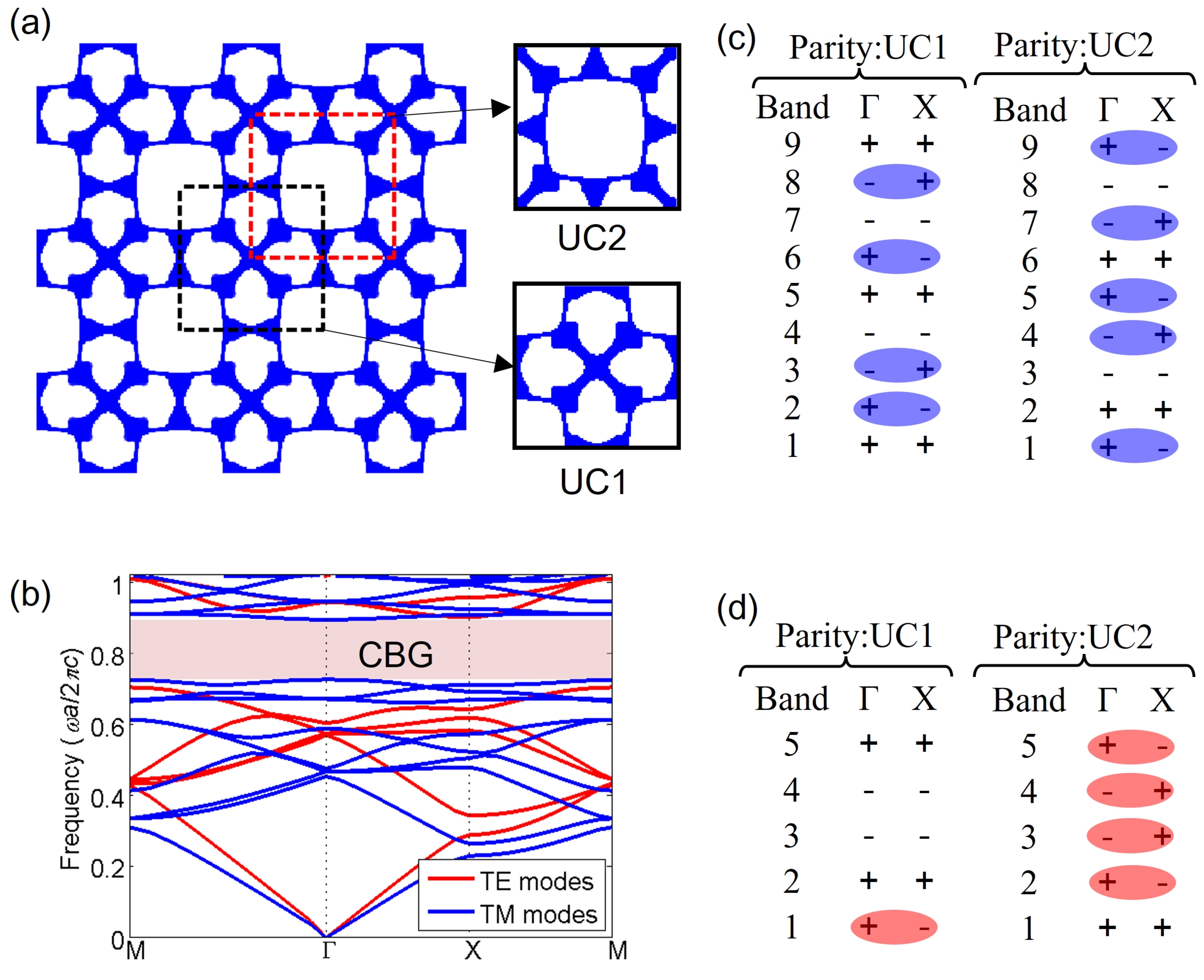} 
\caption{(a) The structure of the optimized PC considered in this work and two different choices of the unit cell, i.e., UC1 and UC2. (b) The band structure of the PC with a CBG. Parities of bands below the CBG at high symmetric points $\Gamma$ and $X$ of the first Brillouin zone for TM (c) and TE (d) modes, respectively, where an odd number of pair of parities with opposite signs at $\Gamma$ and $X$ of the same band below the gap indicates the gap is topological nontrivial whereas an even number implies a trivial gap.}
\label{fig:fig1}
\end{figure}

For 2D photonic crystals (PCs), electromagnetic waves possess two possible polarizations, i.e., transverse magnetic (TM) modes or transverse electric (TE) modes. So far, the existing works of SPTIs have focused on either TM or TE modes. Thus it would be interesting to see whether SPTIs for both the TM and TE modes, i.e., dual-polarization, could be built considering the fact that polarization-dependent manipulation of light is an important and active area of research in photonics~\cite{polarization_review}, which has many practical applications. For example, low-threshold topological nanolasers based on second-order corner states have been demonstrated recently~\cite{Zhang20LightCorner}. A dual-polarization SPTI would allow the realization of the so-called dual-polarization laser~\cite{dp_laser1, dp_laser2, dp_laser3} with additional built-in topological protection. Furthermore, due to the significant difference between the field distributions of TE and TM polarization modes, which response differently to the environment, dual-polarization interferometry has been widely used for a broad range of applications~\cite{dp_Interferometry1, dp_Interferometry2, dp_Interferometry3}, such as, bionanotechnology, surface science, liquid studies,  crystallography and drug discovery.  Dual-polarization is also useful to enhance nonlinear optical effects~\cite{zhang2009ultra, deng2014ultrahigh, dp_nonlinear, minkov2019doubly, Smirnova20APR_NTPreview, Kruk20arxiv_cornerSHG}. For example, 
 in nonlinear optics, materials with large off-diagonal nonlinear susceptibility terms can be used to enhance the harmonic generation when two modes with orthogonal polarizations are mixed~\cite{zhang2009ultra}. Last but not least, polarization-division-multiplexing~\cite{polarization_multiplexing} is widely used  in high speed communication systems due to its ability to enlarge the capacity and data rates of these systems and as such devices, like, dual-polarization modulator~\cite{dp_modulator1, dp_modulator2, dp_modulator3}, transmitter~\cite{dp_transmitter1, dp_transmitter2}, and receiver~\cite{dp_receiver1, dp_receiver2, dp_receiver3} are actively explored for realizing dual-polarization photonic integrated circuits.

Considering the importance of dual-polarization optics and its diverse applications, in this work, we proposes a novel dual-polarization SPTI supporting corner states based on an optimized PC structure with a wide complete band gap (CBG). Interestingly, we find that two different choices of the unit cell (UC) for the optimized PC exhibit distinct topological properties for both the TM and TE modes, where the numerical results are confirmed by simple tight-binding analysis. SPTIs can be built by forming edges and corners between the topologically trivial and nontrivial unit cells. We find that the created SPTI hosts four groups of corner states with different eigenfrequencies and field profiles for both the TM and TE modes.  Moreover, the robustness of these corner states against defects is also verified. The proposed dual-polarization SPTI could provide a promising platform for the development of polarization-independent topological photonic applications. 
   
\section{\label{sec:design}Design of the dual-polarization SPTI}  

Since the corner states generally locate within the frequency band gap, designing a PC with a CBG is the prerequisite for constructing a dual-polarization SPTI. However, dielectric material of PC with band gap in the TM modes usually consists of isolated dielectric ``rods", while it forms ``walls" for the TE modes. This opposite structural characteristics makes it challenging to create a PC with a CBG for the TM and TE modes simultaneously. 
To tackle this design problem, we consider a PC structure composed of GaAs with $\epsilon_r=11.4$ and air with $\epsilon_r=1$ and 
employ the topology optimization method. In this method, we firstly design a PC with a complete band gap for both TM and TE modes and set the gap-midgap ratio as the objective function. We then discretize the unit cell uniformly into $96\times96$ square elements and treat each element as the design variable. 
Upon calculating the sensitivity of the objective function with respect to each design variable, the optimization algorithm increases design variables for elements with high sensitivities and decreases design variables for elements with low sensitivities iteratively until the objective function is maximized (for more details of the numerical optimization algorithm, see Ref. \cite{Meng17ML}). 
Figures \ref{fig:fig1} (a) and (b) show one of the optimized PCs with a maximal CBG and its band structure for both the TM and TE modes, respectively. The gap-midgap ratio of the CBG is 20.79\% with the band gap ranging from the normalized frequency of 0.726 to 0.894. This CBG exists from an overlap of the TM band gap between bands 9 and 10 with the TE band gap between bands 5 and 6. 

For the topological properties of our optimized PC structure, while previous studies based on 2D Su-Schrieffer-Heeger (SSH) model suggest that choosing the unit cell in different ways could result in different topological behaviors \cite{Chen19PRLhighorderTI, Dong19PRLsecondorderTI, Hassan19NPcorner, Ota19Optica, Chen19PRBcornerBIC, Xie18PRBcorner, Liu17PRLzeroBC, Meng20APLacousticHO}, it is a nontrivial question whether our PC will exhibit any topological properties because our system is not based on the SSH design. To study the topological properties of our optimized PC, we select two different UCs (UC1 and UC2) as sketched in Fig. \ref{fig:fig1}(a), where UC1 is the primitive unit cell of the optimized PC while UC2 is a translation of UC1 by $(a/2, a/2)$ with $a$ the lattice constant. The topological properties of UC1 and UC2 could be characterized by the 2D polarization ${\bf P}=(P_x, P_y)$ defined by \cite{Liu17PRLzeroBC} 

\begin{figure}
\includegraphics[width=\columnwidth]{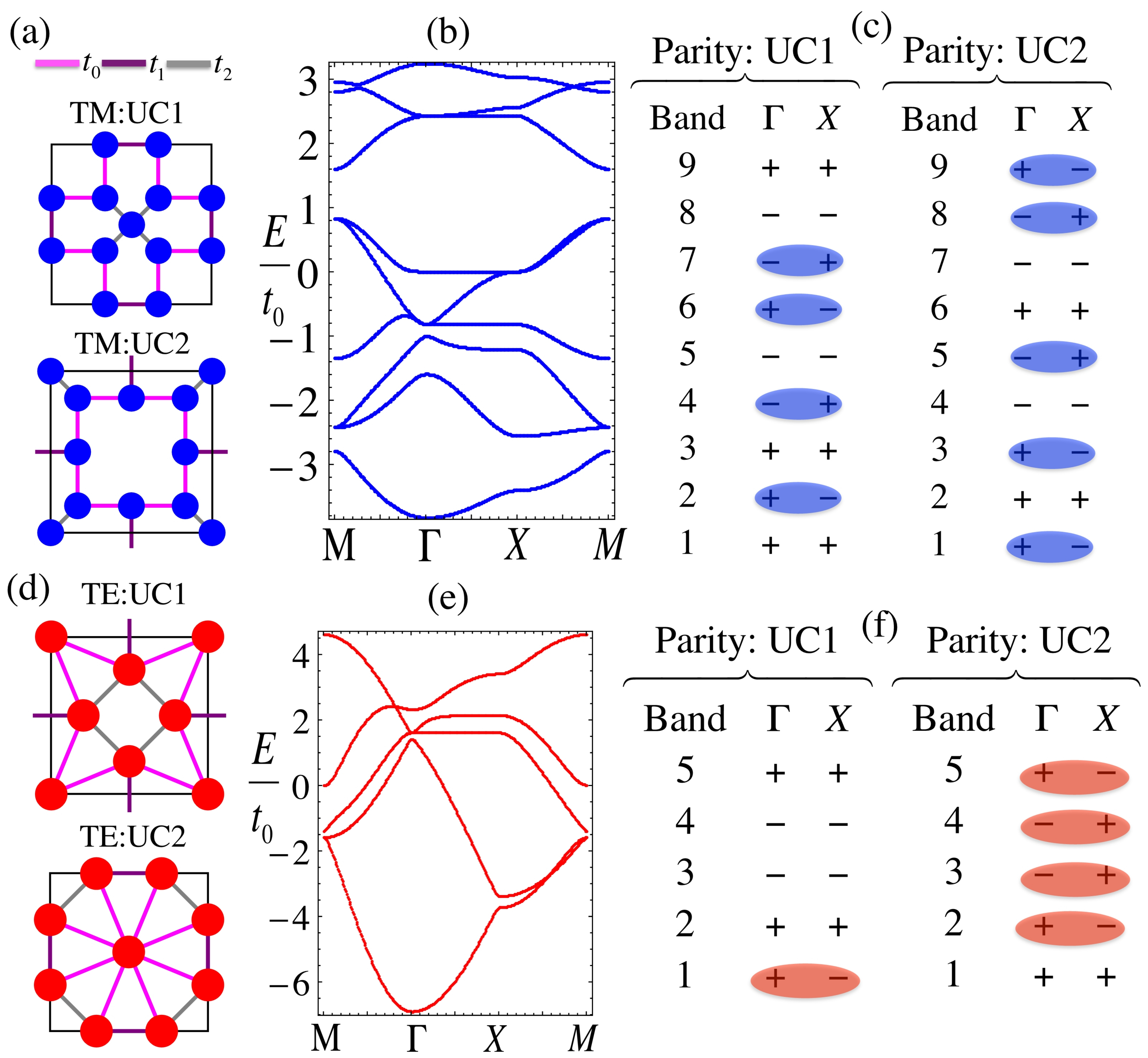} 
\caption{(a) Tight-binding unit cells (UC1, UC2) for the TM modes with three different kinds of hopping $t_0,t_1,t_2$. (b) The tight-binding band structure for the TM modes with $t_0=1, t_1=1.6,t_2=1.4$. (c) Parity distribution of the 9 bands at the high symmetry points of $\Gamma$ and $X$ for the two different unit cells in (a). (d) Tight-binding unit cells (UC1,UC2) for the TE modes. (e) The tight-binding band structure for the TE modes with $t_0=1, t_1=1.6,t_2=1.5$. (f) Parity distribution of the 5 bands at the high symmetry points of $\Gamma$ and $X$ for the two different unit cells in (d). }
\label{fig:fig2}
\end{figure}

\begin{gather}
P_i=\frac{1}{2} \left( \sum_n q_i^n \textrm{modulo}  2 \right), (-1)^{q_i^n}=\frac{\eta_n(X_i)}{\eta_n(\Gamma)}
\label{2Dpol}
\end{gather}
where $i=x,y$ stands for the direction and $\eta_n$ represents the parity at the high symmetry points $\Gamma=(0,0)$ and $X=\pi/a(1,0)$ of the first Brillouin zone for the $n^{th}$ band. Note due to the $C_{4v}$ point group symmetry of our PC structure, ${\bf P}$ could be determined by the parities at the high symmetric points \cite{Liu17PRLzeroBC, Christensen19PRLnonH, Christensen19AM} and also $P_x=P_y$.    

 The parities at the high symmetric points of UC1 and UC2 for both TM and TE modes are shown in  Figs. \ref{fig:fig1}(c) and (d), respectively. The corresponding eigenmode profiles used for identifying the parities are given in Appendix \ref{sec:appendix}, from which one can determine the parity by the behavior of the eigenmode profile under the inversion operation with respect to the center of the unit cell. In this way, one can find that the $p$ modes have an odd parity $(-)$, whereas the $s$ and $d$ modes have an even parity $(+)$. According to the results of Figs. \ref{fig:fig1}(c, d) and Eq. \ref{2Dpol}, which indicate that an odd number of pair of parities with opposite signs at $\Gamma$ and $X$ of the same band below the gap makes the gap topological nontrivial whereas an even number implies the gap is trivial, the 2D polarizations of UC1 and UC2 can be determined to be  ${\bf P}=(0,0) $ and ${\bf P}=(\frac{1}{2},\frac{1}{2})$ respectively for the TM modes, and ${\bf P}=(\frac{1}{2},\frac{1}{2})$ and ${\bf P}=(0,0) $ for the TE modes. This means that UC1 is trivial and UC2 is nontrivial for the TM modes, while UC1 is nontrivial and UC2 is trivial for the TE modes, which is fully compatible with the complementary nature of TM and TE modes. Based on the 2D polarizations, the topological corner charge can be calculated as \cite{Christensen19PRLnonH, Christensen19AM},
 \begin{gather}
Q^c=4P_xP_y
\end{gather}
As a result, the corner charges of (UC1, UC2) are (0, 1) for the TM modes, and (1, 0) for the TE modes. 

\begin{figure*}
\includegraphics[width=\textwidth]{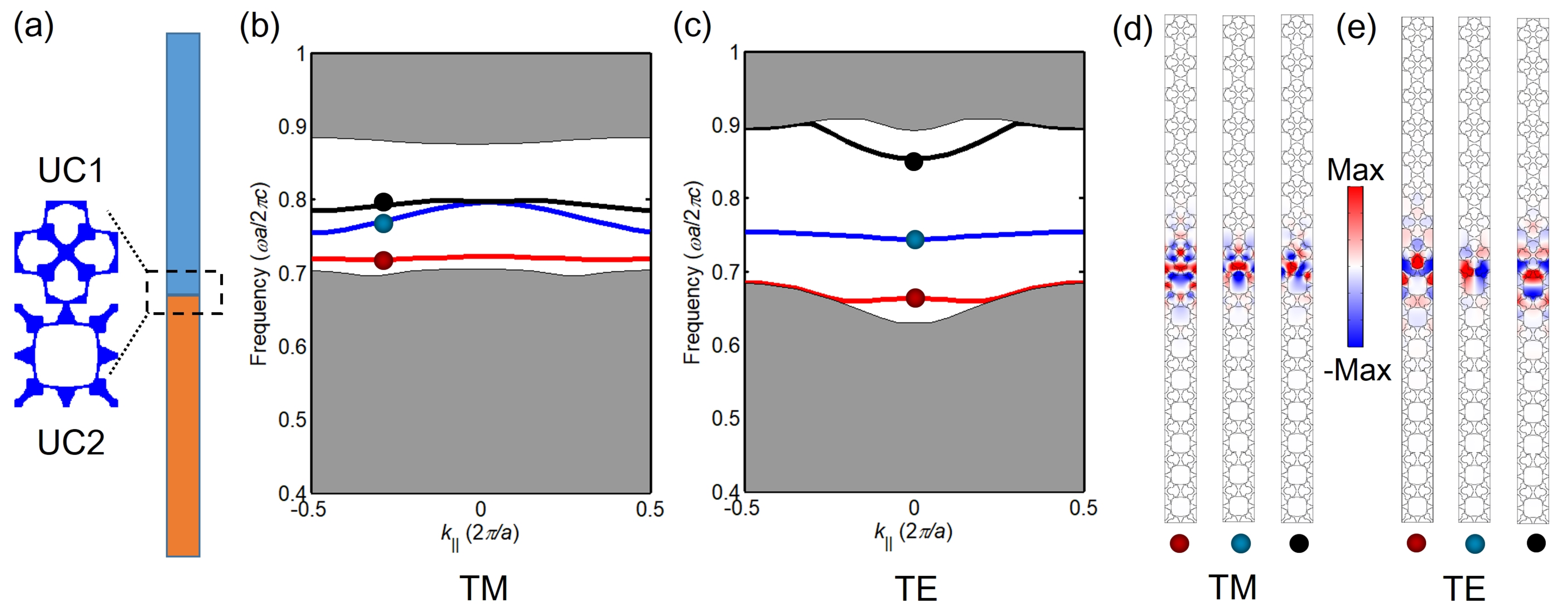} 
\caption{(a) Schematic of the supercell of a ribbon structure which is periodic along the horizontal direction and consists of 8 UC1s and 8 UC2s with a domain wall between them. (b) and (c) show the projected band structures of the ribbon for the TM and TE modes, respectively. The solid lines within the gap indicate the emerging topological edge states while gray regions mark the bulk states.  (d) and (e) show the eigenfield profiles at the solid points in (b) and (c) for TM and TE modes, respectively, which demonstrate the highly localized nature of these modes.}
\label{fig:fig3}
\end{figure*}

While we have shown above that the optimized PC can exhibit topological properties through numerical calculations of the band structure and eigenmode profiles, it would be more insightful if simple tight-binding lattice model could be built to understand these interesting topological features as this would allow us to demonstrate that the topological features of the optimized PC structure exist at a more fundamental level, i.e., applicable to any systems that can effectively implement the lattice model. 

To this end, we note that while the dielectric regions can be viewed as lattice sites for the TM modes, air regions will serve as this propose for the TE modes. As such, the lattice models for the TM and TE modes can readily be constructed, which are presented in Figs. \ref{fig:fig2} (a) and (d). Particles hopping in these lattices could be described by the following Hamiltonian,  
\begin{figure*}
\includegraphics[width=\textwidth]{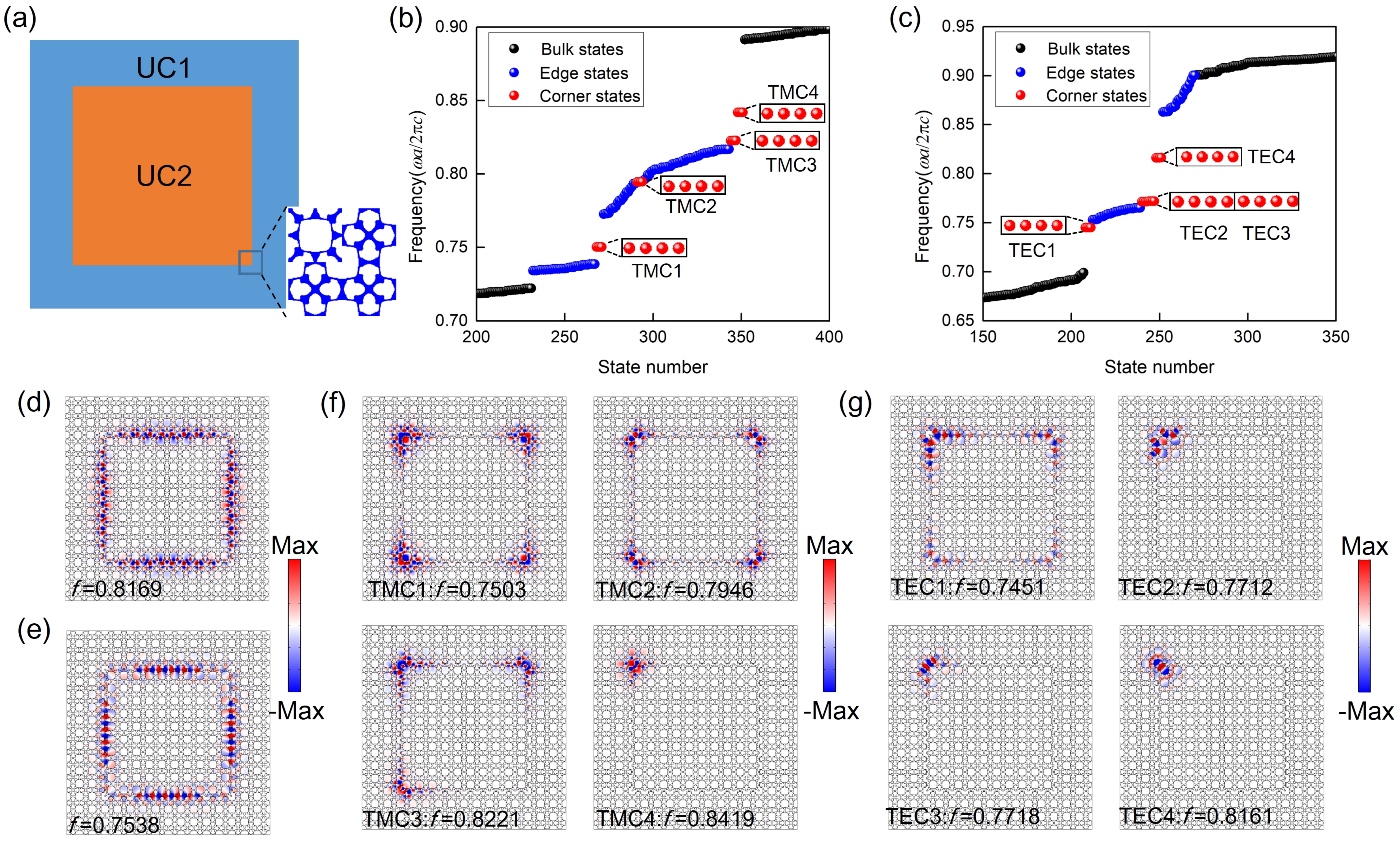} 
\caption{(a) Schematic of a box-shaped finite size structure consisting of $10\times10$  UC2s surrounded by three-layer UC1s for the study of coexistence of topological edge and corner states. (b) and (c) show the calculated eigenfrequencies of the box-shaped structure for the TM and TE modes, respectively, where four groups of corner states emerge within the CBG for both the TM and TE modes. (d) and (e) show the typical eigenfield profiles of the edge states for the TM and TE modes, respectively (note $f=\omega a/2\pi c$). (f) and (g) show the eigenfield distributions of one representative corner state in each of the four groups for the TM and TE modes, respectively.  }
\label{fig:fig4}
\end{figure*}

\begin{gather}
H=-\sum_{\langle ij \rangle}t_{ij} c_i^{\dagger}c_j
\end{gather}
where the hopping patterns ($t_{ij}$) for the TM and TE modes are also given in Figs. \ref{fig:fig2} (a) and (d), respectively and $c_i^{\dagger}$ ($c_i$) is the creation (annihilation) operator of the particle at site $i$.  Note to match more closely with the PC structure in Fig.\ref{fig:fig1}(a), three hopping strengths of $t_0,t_1,t_2$ are introduced.   The values of $t_0,t_1,t_2$ in the tight binding models could be chosen in the following way. First, from the topology-optimized PC structure shown in Fig. \ref{fig:fig1}(a) and its tight-binding pictures in Fig.\ref{fig:fig2}(a) and (d), one can roughly identify the relation among $t_0,t_1,t_2$. For example, for the TM modes, as the dielectric connection for $t_0$ is a thin stripe, it is smaller than $t_1$ and $t_2$. Moreover, one can also see that the dielectric connection for $t_1$ is thicker than $t_2$, thus, $t_1>t_2$. Based on these intuitive relations, one can scan the values of $t_0, t_1, t_2$ in the tight binding models directly and choose these values that give the band structures in Fig.\ref{fig:fig2}(b) and (e) as closer as these in Fig.\ref{fig:fig1}(b). 
The resulting tight-binding band structures for the TM and TE modes are given in Figs.\ref{fig:fig2} (b) and (e) respectively, which reproduce the PC band structures in Fig.\ref{fig:fig1}(b) qualitatively. The lattice structure for the TM modes contains 9 lattice sites in each unit cell while it contains 5 lattice sites for the TE modes, thus there are 9 bands for the TM modes and 5 bands for the TE modes, which perfectly agrees with the number of bands below the CBG for both the TM and TE modes. The lattices in Figs. \ref{fig:fig2} (a) and (d) possess the inversion symmetry, which allows the states at inversion symmetric points (such as $\Gamma$ and $X$) to associate with a definite parity.  Figs. \ref{fig:fig2} (c) and (f) show the parity distributions at the high symmetry points $\Gamma$ and $X$. While the tight-binding band structure and parity distribution for TE modes match these of the PC structure in Fig.\ref{fig:fig1}, we can not find parameters of $t_0,t_1,t_2$ that reproduce the parity distribution of the PC structure for the TM modes as there are more bands for the TM modes than the TE modes, thus requiring a higher degree of fine-tuning. Another possibility is that, the long-range hopping physics of the electromagnetic waves in the PC structure is not captured by our simple tight-binding description with only three short-range hopping terms. Nevertheless, the topological invariants (odd or even number of pairs of parity with opposite signs at $\Gamma$ and $X$ for the same band) for the TM modes are the same as these of the PC structure. 

It is interesting to note that our lattice models have odd number of bands (i.e., 9 bands for the TM modes and 5 for the TE) and the whole band structure (i.e.,considered gap above all bands) possesses topological behavior for one choice of the unit cell, which is very different from the widely used 2D SSH square tight-binding lattice model (see, e.g., Ref. \cite{Liu17PRLzeroBC}), which has even number of bands and the whole band structure is topological trivial for both choices of the unit cell, i.e., one would need to consider partially filled bands of the 2D square SSH model for topological applications.


\section{\label{sec:edge-corner} Topological edge and corner states }

According to the bulk-edge correspondence principle, topological edge states will emerge at the boundary between topological trivial and nontrivial PCs. To verify the existence of topological edge states, we consider a ribbon structure consisting of 8 UC1s and 8 UC2s with a domain wall between them, as illustrated in Fig. \ref{fig:fig3}(a). Figures \ref{fig:fig3}(b) and (c) show the projected band structures for the TM and TE modes, respectively, from which one can see that three waveguide modes appear within the CBG. Figures \ref{fig:fig3} (d) and (e) further present the eigenfields of the edge states calculated at the solid points in Figs. \ref{fig:fig3} (b) and (c) for the TM and TE modes, respectively. As can be seen, these modes are highly localized at the edge between UC1 and UC2, which could be used as defect-immune light transport.  Importantly, these edge states are gapped, which is also the prerequisite for the generation of topological corner states \cite{Christensen19PRLnonH, Christensen19AM}. 

\begin{figure*}
\includegraphics[width=\textwidth]{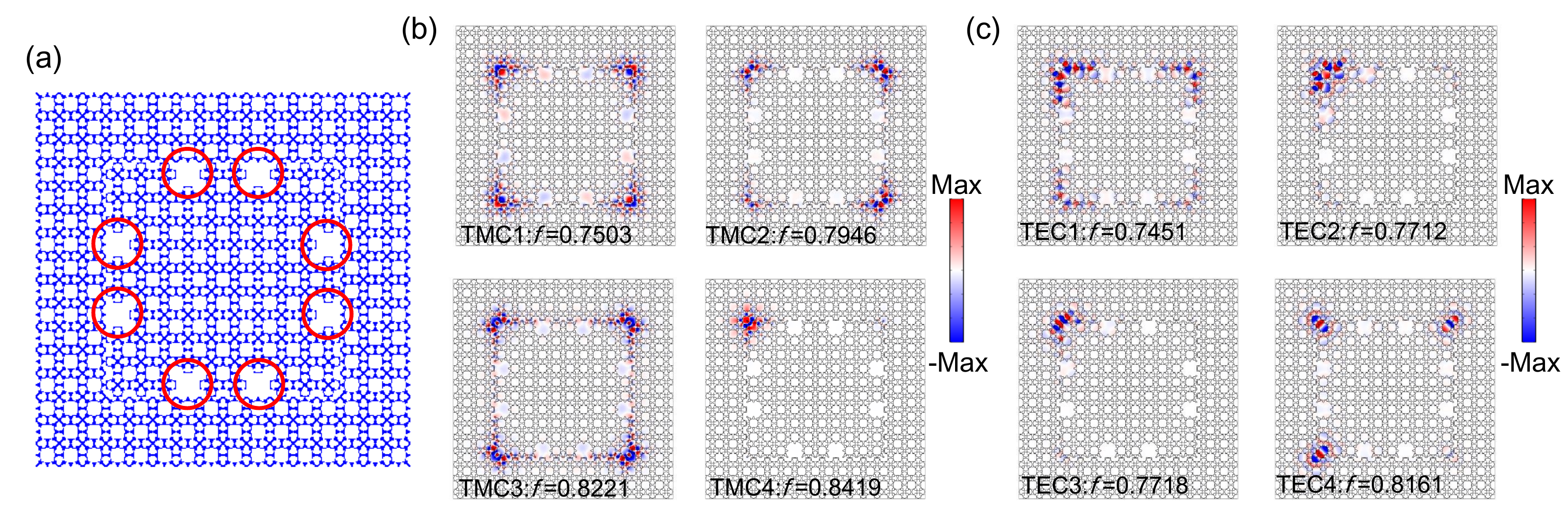} 
\caption{(a) Schematic of a perturbed structure after removing certain UCs donated by the red circles for the demonstration of robustness of the corner states agains defect. (b) and (c) show the calculated eigenfields and eigenfrequencies of one representative corner state from each of the four groups for the TM and TE modes, respectively.}
\label{fig:fig5}
\end{figure*}

The different corner charges between UC1 and UC2 guarantee the existence of topological corner states. To further demonstrate the coexistence of topological edge and corner states, we build a box-shaped region consisting of $10\times10$ UC2s surrounded by three-layer UC1s, as sketched in Fig. \ref{fig:fig4}(a). Figures \ref{fig:fig4}(b) and (c) plot the calculated eigenfrequencies of the finite size structure for TM and TE modes, respectively, from which one can see the coexistence of topological edge and corner states within the CBG for both the TM and TE modes. Figures \ref{fig:fig4}(d) and (e) show the typical eigenfield distributions of the edge states for the TM and TE modes, respectively. It can be seen that electromagnetic waves are highly localized at the domain wall between UC1 and UC2. It is interesting to note that the proposed SPTI hosts four groups of corner states at different frequencies, which are labelled as TMC1, TMC2, TMC3, TMC4 for the TM modes and TEC1, TEC2, TEC3, TEC4 for the TE modes in Figs. \ref{fig:fig4}(b) and (c) respectively. As a result, our current design of the SPTI is distinct from any existing SPTIs which only host one or two groups of corner states. Figure \ref{fig:fig4}(f) and (g) show the eigenfield distributions of the corner states for TM and TE modes, respectively. Each group of corner states contains four degenerate corner states, therefor only one eigenfield distribution of a representative corner state in each group is plotted. It reveals that these corner states are highly localized around the corners and exhibit different profiles, which may provide flexibilities in practical applications.

The corner states have a topological origin related to the different corner charges of UC1 and UC2, thus they enjoy the topological protection. To demonstrate the robustness of the corner states against defects, we remove some unit cells around the corners as denoted by the red circles in Fig. \ref{fig:fig5} (a). The eignfields and eigenfrequencies of the corner states in this perturbed structure are calculated and presented in Figs. \ref{fig:fig5} (b) and (c) for  the TM and TE modes, respectively. As can be seen, although defects are introduced around the corners, all the corner states remain unchanged and their eigenfrequencies also keep invariant, demonstrating the robustness of the corner states against defects.


\section{\label{sec:conclusion} Conclusion}

In conclusion, we have demonstrated the design of a dual-polarization SPTI, which supports localized corner states for both the TM and TE modes in a wide common gap. Our design is based on the topology optimization method and the topological properties of the optimized PC structure are characterized through the band structures and parity analysis of the eigenmode profiles. The designed SPTI hosts four groups of corner states with different eigenfrequencies and field profiles for both the TM and TE modes, which provide more degrees of freedom for developing new photonic devices compared to the existing SPTIs. We have also demonstrated the robustness of these corner states agains defects, which could advance the design of polarization-independent topological devices with diverse functionalities. 

Moreover, we have constructed simple tight-binding lattice models, which allow us to capture the key topological features of the optimized PC structure in a transparent way and these tight-binding lattice models indicate that the dual-polarization topological physics we have revealed in the PC system carries over to diverse system platforms, such as plasmonics, acoustics, as long as these systems could implement the tight binding models. Furthermore, these tight-binding lattice models have odd number of lattice sites in each unit cell and the whole band structure shows topological properties for one choice of the unit cell. These interesting features are beyond the 2D square SSH paradigm, which has even number of lattice sites in its unit cell and the whole band structure shows no topological properties for both choices of the unit cell, i.e., one would need to consider partially filled bands of the 2D square SSH model for topological applications, which may limit the size of the topological band gap. Thus we expect our findings would bring new insights and open interesting directions in this field. Finally, we would like to note that it is certainly interesting to see whether dual-polarization SPTI could be constructed in PCs having hexagonal symmetry \cite{Liu19PRLquadrupole, Christensen20PRBsonic, Corner_QSH_NC20}.

\begin{figure*}
\includegraphics[width=\textwidth]{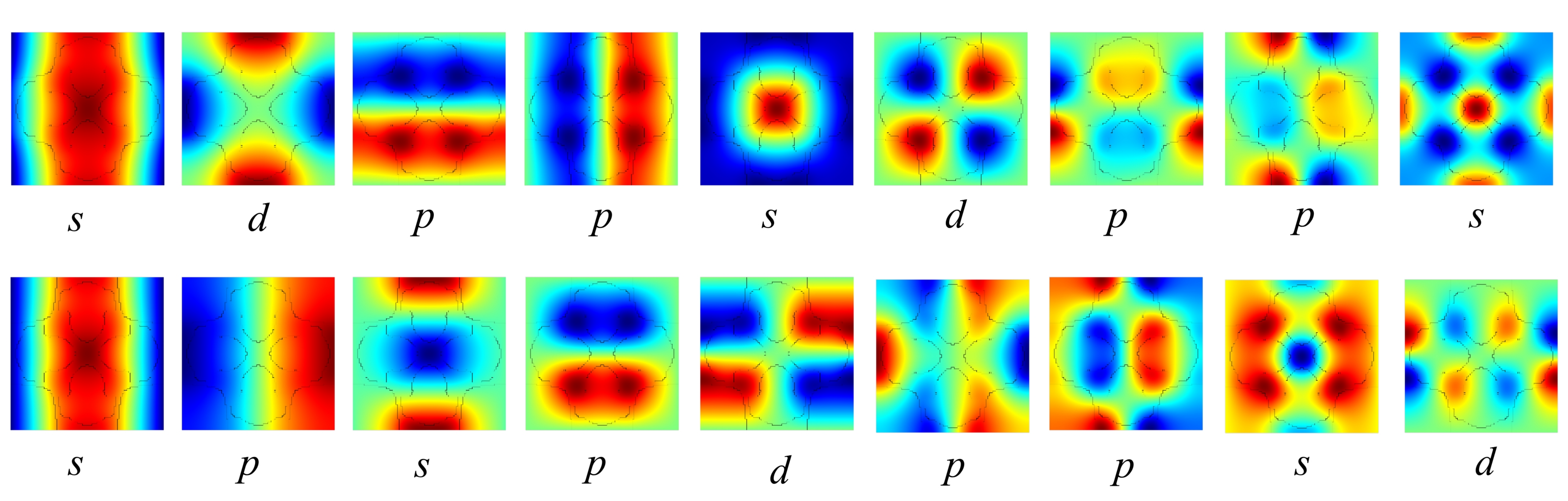} 
\caption{Eigenmode profiles at $\Gamma$ point (upper panel) and X point (bottom panel) for the nine bands (from left to right) of UC1 in TM modes. }
\label{fig:fig6}
\end{figure*}

\begin{figure*}
\includegraphics[width=\textwidth]{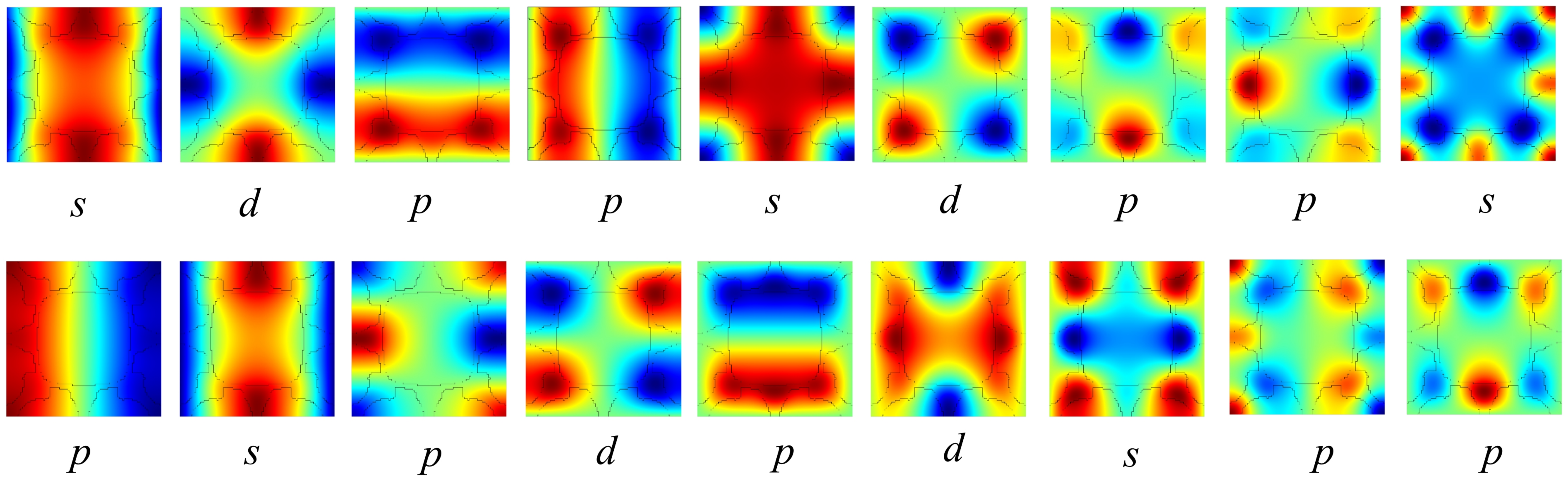} 
\caption{Eigenmode profiles at $\Gamma$ point (upper panel) and X point (bottom panel) for the nine bands (from left to right) of UC2 in TM modes. }
\label{fig:fig7}
\end{figure*}

\begin{figure*}
\includegraphics[width=0.6\textwidth]{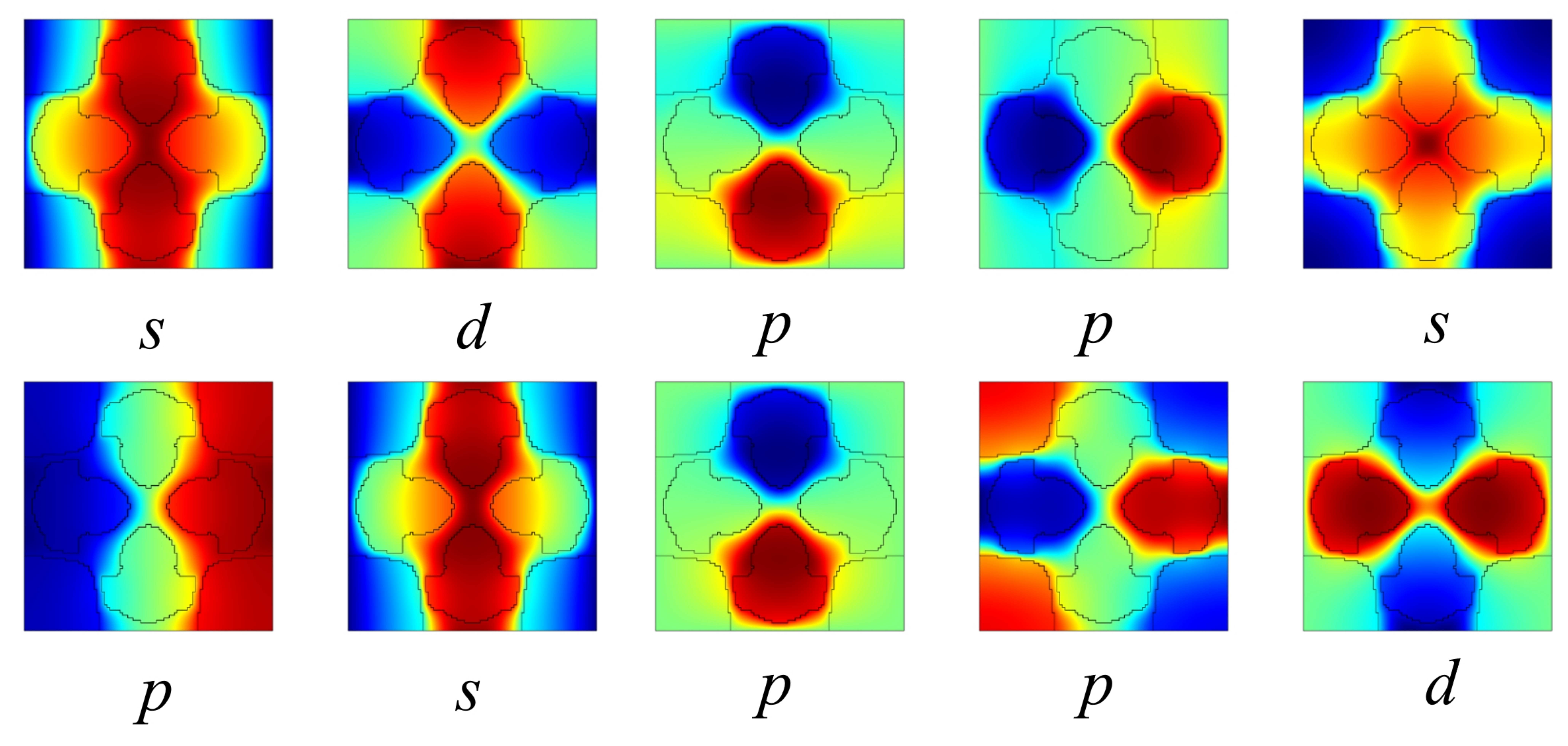} 
\caption{Eigenmode profiles at $\Gamma$ point (upper panel) and X point (bottom panel) for the five bands (from left to right) of UC1 in TE modes. }
\label{fig:fig8}
\end{figure*}

\begin{figure*}
\includegraphics[width=0.6\textwidth]{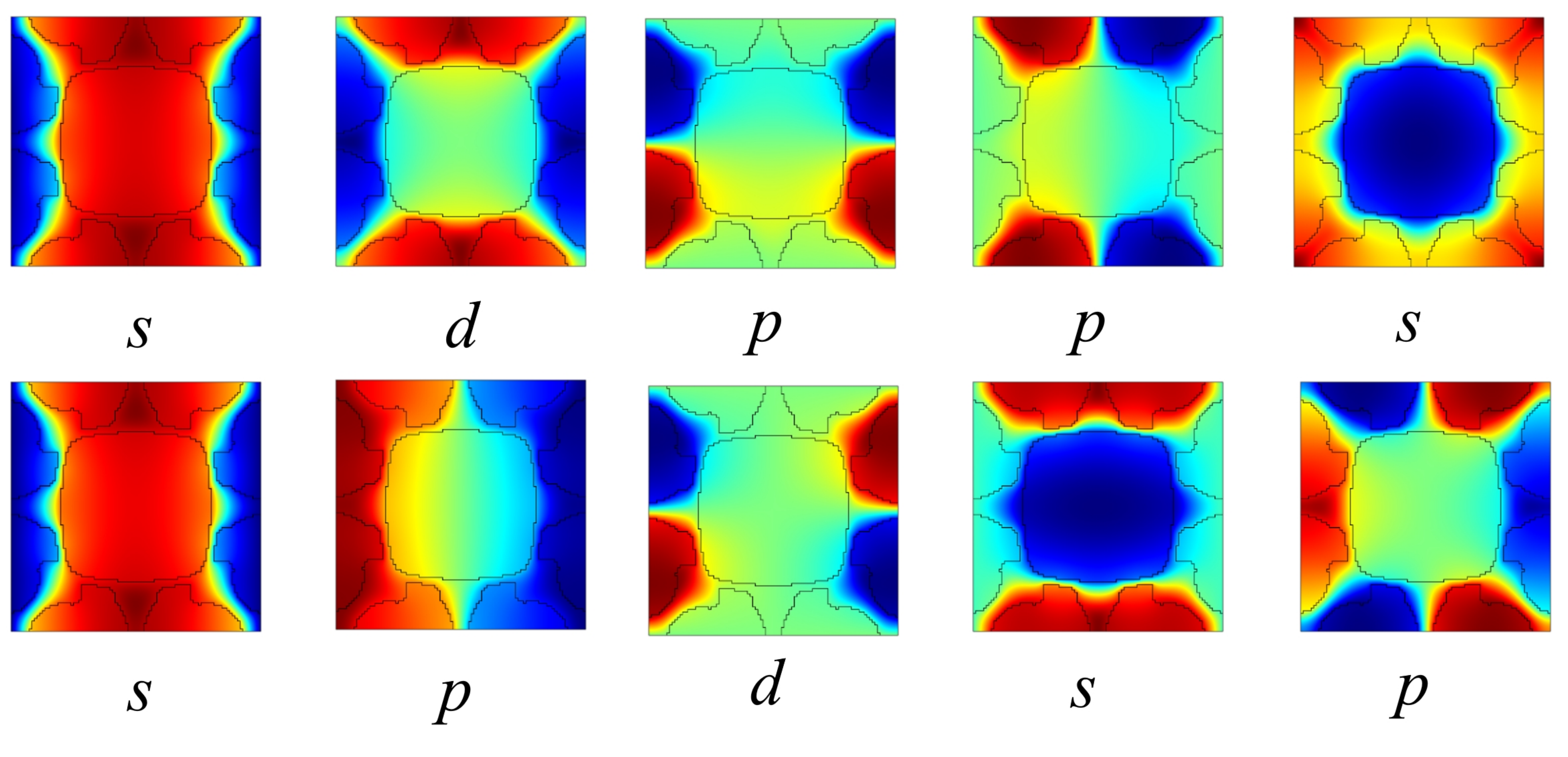} 
\caption{Eigenmode profiles at $\Gamma$ point (upper panel) and X point (bottom panel) for the five bands (from left to right) of UC2 in TE modes. }
\label{fig:fig9}
\end{figure*}

\appendix
\section{\label{sec:appendix}Eigenmode profiles of the optimized PC at high symmetry points $\Gamma$ and $X$}

In this appendix, we give the eigenmode profiles of the optimized PC at high symmetry points of $\Gamma$ and $X$, whose parities under the inversion operation are listed in Fig.\ref{fig:fig1} (c) and (d) and used for determination of the  topological properties of UC1 and UC2 through Eq. \ref{2Dpol}. In specific, Figs. \ref{fig:fig6} and \ref{fig:fig7} show the TM eigenmode profiles of the 9 bands at the high symmetry points of $\Gamma$ and $X$ for UC1 and UC2 respectively. On the other hand, Figs. \ref{fig:fig8} and \ref{fig:fig9} show the TE eigenmode profiles of the 5 bands at the high symmetry points of $\Gamma$ and $X$ for UC1 and UC2 respectively. Note, for the eigenmode profiles presented in Figs. \ref{fig:fig6} \ref{fig:fig7} \ref{fig:fig8} and \ref{fig:fig9}, while the $p$ modes have an odd parity $(-)$, the $s$ and $d$ modes have an even parity $(+)$. 


\begin{acknowledgments}
The authors thank Professor Yuri Kivshar at the Nonlinear Physics Center, Australian National University for valuable feedback. The authors also wish to acknowledge the support from the Hong Kong Scholars Program (Grant No. XJ2020004), the Australian Research Council through the Discovery Project schemes (DP210103523; DP190103186) and the Industrial Transformation Training Centres scheme (IC180100005).
\end{acknowledgments}

\end{document}